\documentclass[aps, prd, nofootinbib, preprint]{revtex4-1}
\usepackage{graphicx,graphics}

\usepackage{dcolumn}
\usepackage{amsmath}
\usepackage{array}
\usepackage{bm}
\usepackage{amssymb}
\usepackage{amsfonts}
\usepackage{color}

\begin{document}

\title{Glueballs and the Yang-Mills plasma in a $T$-matrix approach}

\author{Gwendolyn \surname{Lacroix}, Claude \surname{Semay}}
\email[E-mail: ]{gwendolyn.lacroix@umons.ac.be}
\email[E-mail: ]{claude.semay@umons.ac.be}
\affiliation{Service de Physique Nucl\'{e}aire et Subnucl\'eaire,
Universit\'{e} de Mons, 
UMONS Research Institute for Complex Systems,
Place du Parc 20, 7000 Mons, Belgium}

\author{Daniel \surname{Cabrera}}
\email[E-mail: ]{Daniel.Cabrera@fis.ucm.es}
\affiliation{Departamento de F{\'i}sica Te{\'o}rica II, Universidad Complutense, 28040 Madrid, Spain.}

\author{Fabien \surname{Buisseret}}
\email[E-mail: ]{fabien.buisseret@umons.ac.be}
\affiliation{Service de Physique Nucl\'{e}aire et Subnucl\'{e}aire,
Universit\'{e} de Mons -- UMONS,
Place du Parc 20, 7000 Mons, Belgium;\\ 
Haute \' Ecole Louvain en Hainaut (HELHa), Chauss\'ee de Binche 159, 7000 Mons, Belgium}

\date{\today}

\begin{abstract}
The strongly coupled phase of Yang-Mills plasma with arbitrary gauge group is studied in a $T$-matrix approach. The existence of lowest-lying glueballs, interpreted as bound states of two
transverse gluons (quasi-particles in a many-body set up), is analyzed in a non-perturbative scattering formalism with the input of lattice-QCD static potentials. Glueballs are actually found to be bound up to 1.3 $T_c$. Starting from the $T$-matrix, the plasma equation of state is computed by resorting to Dashen, Ma and Bernstein's formulation of statistical mechanics and favorably compared to quenched lattice data. Special emphasis is put on SU($N$) gauge groups, for which analytical results can be obtained in the large-$N$ limit, and predictions for a $G_2$ gauge group are also given within this work.
\end{abstract}

\pacs{12.38.Mh, 12.39.Mk, 11.15.Pg}

\maketitle

\section{Introduction}

More than two decades after pioneering works \cite{coll, shur}, the phenomenology related to the deconfined phase of QCD, \textit{i.e.}~the quark-gluon plasma (QGP) is still a fascinating topic both experimentally and theoretically. On the experimental side, the QCD matter was or is studied in heavy-ion collisions (RHIC, SPS, FAIR, LHC). These experiments seem to show  that the QGP behaves like a perfect fluid. On the theoretical side, the study of QCD at finite temperature deserves also much interest because it is a challenging problem in itself and because of the many connections with experiments. 

\par The aim of this work is to study the thermodynamic features of QGP by resorting to a $T$-matrix approach. The power of this approach is that the bound states and scattering states of the system can be studied in a whole picture. Such an approach has already proved to give relevant results in the study of hadronic matter above the critical temperature of deconfinement ($T_c$) \cite{cabre06} but has not yet been applied to compute the equation of state (EoS). This  observable will be performed here thanks to the Dashen, Ma and Bernstein's formulation of statistical mechanics in terms of the ${S}$-matrix (or ${T}$-matrix)~\cite{dashen}. Such a formulation is particularly well suited for systems whose microscopic constituents behave according to relativistic quantum mechanics. The QGP is indeed identified to a quantum gas of gluons and quarks, which are seen as the effective degrees of freedom propagating in the plasma. This assumption is actually common to all the so-called quasiparticle approaches \cite{quasip}, with the crucial difference that the use of a ${T}$-matrix  formulation allows us to investigate the behavior of the QGP in a temperature range where it is strongly interacting. This strong interaction means here that bound states are expected to still survive above $T_c$. 

\par Although the above formulation can be applied to the full QGP, this paper is dedicated to the description of the gluon plasma. Dealing with only one particle species simplifies drastically the problem while the main feature of the description, \textit{i.e.}~the explicit inclusion of interactions in a quasiparticle approach, is kept. Moreover, the pure gauge thermodynamic features (in particular, the EoS) are well-known in lattice QCD; This will allow an accurate comparison  between our phenomenological approach and the lattice QCD calculations. 

\par A particularity of this paper is the generalization of the formalism to any gauge groups, with a particular attention for SU($N$) and the large-$N$ limit, and for G$_2$. This group has originally attracted attention because, the center of G$_2$ being trivial, models relating deconfinement to the breaking of a center of symmetry are no longer valid as for SU($N$). However, it still exhibits a first-order phase transition as SU($N$) does \cite{pepe}. Hence, G$_2$ appears quite attractive from a theoretical point of view. 

\par The paper is organized as follows. Sec.~II is dedicated to the presentation of the general quasiparticle approach based on the ${T}$-matrix formalism proposed in~\cite{dashen}. In Sec.~III, the model is particularized to a Yang-Mills plasma with the inclusion of 2-body interactions and, in Sec.~IV, useful analytic comments concerning the thermodynamic observables in the SU($N$) and G$_2$ cases are discussed. The model parameters are fixed in Sec.~V and the existence of the bound states inside the gluon plasma is discussed in Sec.~VI. In Sec.~VII, the computation of the EoS is presented. Finally, Sec.~VIII is devoted to the conclusions and perspectives.

\section{$T$-matrix formalism}\label{Tmatsec}
\subsection{Generalities}

The results of~\cite{dashen} can be summarized as follows: The grand potential $\Omega$, expressed as an energy density, of an interacting particle gas is given by (in units where $\hbar=c=k_B=1$).
\begin{equation}\label{pot0}
\Omega=\Omega_0+\sum_\nu\left[\Omega_\nu-\frac{{\rm e}^{\beta\vec \mu\cdot\vec  N}}{2\pi^2\beta^2}\int^\infty_{M_\nu} \frac{d\epsilon}{4\pi i}\, \epsilon^2\,  K_2(\beta\epsilon)\, \left. {\rm Tr}_\nu \left({\cal S}S^{-1}\overleftrightarrow{\partial_\epsilon}S \right)\right|_c\right].
\end{equation}  
In the above equation, the first term, $\Omega_0$, is the grand potential of the free relativistic particles, \textit{i.e.} the remaining part of the grand potential if the interactions are turned off. The second term accounts for interactions in the plasma and is a sum running on all the species, the number of particles included, and the quantum numbers necessary to fix a channel. The set of all these channels is generically denoted $\nu$. The vectors $\vec \mu=(\mu_1, \mu_2,\dots)$ and $\vec N=(N_1,N_2,\dots)$ contain the chemical potentials and the particle number of each species taking part in a given scattering channel. 

The contributions above and below the threshold\footnote{Within this approach, the threshold is the summation on the mass of all the particles included in a given channel $\nu$.} $M_\nu$ are separated. Below the threshold, one has $\Omega_\nu$ the grand potential coming from bound states, seen as free additional species in the plasma and appearing as poles of the ${S}$-matrix. Above the threshold, one has the scattering contribution, where the trace is taken in the center of mass frame of the channel $\nu$ and where $S$ is the ${S}$-matrix, depending in particular on the total energy $\epsilon$. The symmetrizer ${\cal S}$ enforces the Pauli principle when a channel involving identical particles is considered, and the subscript $c$ means that only the connected scattering diagrams are taken into account. Notice that $K_2(x)$ is the modified Bessel function of the second kind, that $\beta$ is linked to the temperature $T$ thanks to $\beta = 1/T$, and that the notation $A\overleftrightarrow{\partial_x} B=A(\partial_xB)-(\partial_xA)B$ is used.

By definition, $S=1-2\pi i\, \delta(\epsilon-H_0)\, {\cal T}$, where ${\cal T}$ is the off-shell $T$-matrix and where $H_0$ is the free Hamiltonian of the system. A convenient way to compute ${\cal T}$ is to solve the Lippmann-Schwinger equation for the off-shell $T$-matrix, schematically given by
\begin{equation}\label{ls}
{\cal T}=V+ V\, G_0\, {\cal T},
\end{equation}
with $G_0$ the free propagator and $V$ the interaction potential. 

Once the ${T}$-matrix is known, two problems can be simultaneously addressed:
The existence of bound states in the plasma and its EoS. The $T$-matrix
formalism has the advantage of treating bound and scattering states on the same
footing, and is particularly suited for the present situation where we expect
bound states to become less and less bound when the temperature increases,
eventually crossing over and melting into the continuum. This dissociation
mechanism has been shown to provide considerable threshold enhancement effects
in heavy  quark anti-quark correlation functions \cite{cabre06}.

The plasma EoS is obtained from (\ref{pot0}). Then, the pressure is simply given by 
\begin{equation}
p=-\Omega ,
\end{equation}
and the other thermodynamic observables can derived from $p$. For example, the trace anomaly ($\Delta=e-3\,p$) and the entropy density ($s$) read
\begin{equation}
\Delta=-\frac{1}{\beta^3} \left[ \partial_\beta\left(\beta^4 p\right)\right] _{{\cal V} \text{,}\, \beta\vec \mu}, \quad s=-\beta^2\left[ \partial_\beta p\right]_{{\cal V} \text{,}\, \vec \mu}
\end{equation}
where ${\cal V}$ is the volume of the system.
For later convenience, the thermodynamic quantities will be normalized to the Stefan-Boltzmann pressure, which is defined as
\begin{equation}\label{psb}
p_{SB}=-\lim_{m_i\rightarrow 0}\Omega_0,
\end{equation}
$m_i$ being the masses of the particles propagating in the medium. 

\subsection{Interaction potential}

The explicit computation of $\Omega$ obviously requires the knowledge of the on-shell $T$-matrix that can be derived in particular from (\ref{ls}). A key ingredient of the present approach is thus the potential $V$, encoding the interactions between the particles in the plasma. In the following, $V$ is chosen as pairwise: For a $n$-body channel, $V=\sum_{i<j} V_{ij}$ with $V_{ij}$ the potential between two particle species $i$ and $j$. Having in mind the building of an effective framework describing the deconfined phase of a non-abelian gauge theory, each particle composing the plasma should be in a given representation of the considered gauge (or color) group. It is therefore reasonable to assume that the potential $V$ between two particles in the representations $R_i$ and $R_j$ of the considered gauge group has the color-dependence of a (screened) one-gluon-exchange potential, that is, in momentum space,
\begin{equation}\label{V00}
V_{ij}=\tilde M_{R_i} \cdot \tilde M_{R_j} \alpha_S\, \bar v(\beta,\vec{q},\vec{q}\,'),
\end{equation}
where $\tilde M_{R}$ denotes the generator of the considered gauge algebra in the representation $R$, and where the real function $\bar v$ only depends on the temperature and two momenta (no dependence on the mass or other attributes of the particle). We keep the name gluon for the gauge particle even if the gauge group can formally be arbitrary. In the above definition, it has to be remembered that $\alpha_S=g^2/4\pi$ and that $g^2=\lambda/C_2^{ adj}$, $adj$ being the adjoint representation of the gauge group under study and $C_2^R$ being the value of the quadratic Casimir in the representation $R$.  Note that in the case of SU($N$), $\lambda$ is the 't~Hooft coupling (fixed in the large-$N$ limit). 

Introducing quadratic Casimirs, one can rewrite (\ref{V00}) as  
\begin{equation}\label{V0}
V_{ij}=\frac{C^{{\cal C}}_2-C^{R_i}_2-C^{R_j}_2}{2}\, \alpha_S\, \bar v\equiv \kappa_{{\cal C};ij}\,  v,
\end{equation}
with ${\cal C}$ the pair representation and
\begin{equation}\label{kappa0}
\kappa_{{\cal C};ij}=\frac{C^{{\cal C}}_2-C^{R_i}_2-C^{R_j}_2}{2\,C_2^{ adj}}.
\end{equation}
Again, the real function $v=v(\beta,\vec{q},\vec{q} \,')$ only depends on the temperature and on two momenta --  an explicit form for $v$ will be given later. The validity of the form (\ref{V0}) for $V_{ij}$ has partially been checked in pure gauge SU(3) lattice calculations, showing that the static potential between two sources, in different representations and bound in a color singlet, follows the Casimir scaling expected from a process of one-gluon-exchange type~\cite{gupta}. The peculiar scaling (\ref{V0}) also leads to a relevant large-$N$ behavior of the EoS when the gauge group SU($N$) is chosen, as it will be shown in Sec.~\ref{sunT}. 

Among the various possible representations, the case where a singlet (denoted $\bullet$) appears in the tensor product $R_i\otimes R_j$ is particularly relevant: Since $C_2^\bullet=0$ and the other quadratic Casimirs are positive, the singlet is the most attractive channel in any two-body scattering process, so the most favorable one for the formation of bound states. Such two-particle bound states should presumably be the lowest-lying ones and, being color-singlet, would give rise to low-lying glueballs or mesons for instance. 

\subsection{Born approximation} \label{Born}

The scattering term in (\ref{pot0}), given by
\begin{equation}\label{pot_s}
\Omega_s= - \sum_{\nu}\frac{{\rm e}^{\beta\vec\mu\cdot\vec N}}{2\pi^2\beta^2} \int^\infty_{M_\nu} \frac{d\epsilon}{4\pi i}\, \epsilon^2\,  K_2(\beta\epsilon)\, \left. {\rm Tr}_\nu \left({\cal S}S^{-1}\overleftrightarrow{\partial_\epsilon}S \right)\right|_c ,
\end{equation}  
can be considerably simplified by using the Born approximation, \textit{i.e.} by noticing that if the interactions are weak enough, ${\cal T}=V+{\rm O}(V^2)$. Such conditions are generally expected to be valid at high enough temperatures, where the typical interaction energy is small with respect to the typical thermal energy of the particles. Note also that, in some cases, this approximation can be relevant when the factor $\kappa_{{\cal C};ij}$ is negligible, irrespective of the temperature: Such cases will be encountered when the gauge group is SU($N$) (see Sec.~\ref{sunT}). 
\par To the first order in $V$, (\ref{pot_s}) becomes
\begin{equation}\label{pot_s2}
\Omega_s =  \sum_\nu \frac{{\rm e}^{\beta\vec\mu\cdot\vec N}}{2\pi^2\beta^2}\int^\infty_{M_\nu} d\epsilon \, \epsilon^2\,  K_2(\beta\epsilon)\, \left. {\rm Tr}_\nu \,\partial_\epsilon (\delta(\epsilon -H_0) V )\right|_c + {\rm O}(V^2).
\end{equation}  
Let us write explicitly $\nu=(n,\tilde \nu)$, where $n$ is the total number of particles involved in a given scattering channel, and where $\tilde \nu$ are the remaining quantum numbers. A useful remark to be done at this stage is that the pairwise structure of $V$ causes $ \left.V \right|_c$ to be always vanishing excepted in two-body channels. Here, at the Born approximation, $n$ is always equal to 2. Then,  $\left. {\rm Tr}_\nu \, \partial_\epsilon (\delta(\epsilon -H_0) V) \right|_c= {\rm Tr}_{\tilde \nu} \,\kappa_{{\cal C},ij}\,\partial_\epsilon (\delta v)$, with $\delta=\delta(\epsilon-\epsilon_{ij}(q))$ and $\epsilon_{ij}(q)=\sqrt{q^2+m_i^2}+\sqrt{q^2+m_j^2}$. Note that the color channel ${\cal C}$ and the particles species  $i$, $j$ are part of $\tilde \nu$. After having extracted from the trace the color and $J^{PC}$ dependences ($J^P$ if the charge conjugation is not relevant), one is led to 
\begin{equation}\label{pot_s3}
\Omega_s= \sum_{(i,j)}\frac{{\rm e}^{\beta(\mu_{i} + \mu_j)}}{2\pi^2\beta^2}\sum_{J^{PC}} (2J+1) \sum_{{\cal C}} {\rm dim}\,{\cal C} \, \kappa_{{\cal C},ij}   \int^\infty_{M_{\tilde\nu}} d\epsilon \, \epsilon^2\,  K_2(\beta\epsilon)\,  {\rm Tr}_q \,\partial_\epsilon (\delta \, v_{J^{PC}} ) +{\rm O}(V^2),
\end{equation}
with ${\rm dim}\,{\cal C}$ is the pair representation dimension, ${\rm Tr}_{q}$ the remaining trace on the momentum space and $v_{J^{PC}}$ the potential with the angular symmetry of the considered channel. 
\par Among the various summations to be performed in $\sum_{\tilde \nu}$, two are of particular interest: The one over the different interacting species, that can be denoted $\sum_{(i,j)}$, and the one over the representations appearing in $R_i\otimes R_j$, that is $\sum_{{\cal C}}$. Because of $\kappa_{{\cal C},ij}$, (\ref{pot_s3}) is thus proportional to a factor $\sum_{{\cal C}}{\rm dim}\,{\cal C} \ \kappa_{{\cal C},ij}$ for a given pair $i$, $j$ in a given $J^{PC}$ channel. When the combinations of species does not have to respect a symmetry principle, this last sum runs on all the representations appearing in $R_i\otimes R_j$; one can then show that  
\begin{equation}\label{group}
\sum_{{\cal C}}{\rm dim}\,{\cal C} \ \kappa_{{\cal C},ij}=0.
\end{equation}
Indeed, it is known in group theory that the second order Dynkin indices $I^ R$ in a tensor product obey a sum rule that can be rewritten using our notations as $I^{R_i}\, {\rm dim}\,R_j+I^{R_j}\, {\rm dim}\,R_i=\sum_{{\cal C}} I^{{\cal C}}$ \cite{group}. Using $C_2^{R}=({\dim }\,{adj} /{\dim }\,R) I^R$ \cite{group}, one straightforwardly shows that (\ref{group}) holds. Note that (\ref{group}) and (\ref{pot_s3}) are thus \textit{a priori} nonzero when a symmetry principle has to be respected: The summation cannot then be performed on all possible color representations.

\section{Yang-Mills plasma}\label{YM}

\subsection{Grand canonical potential}

Let us now particularize the general formalism presented in the previous section to a genuine Yang-Mills plasma, \textit{i.e.}~with no matter fields. The bosonic degrees of freedom propagating in the plasma are then quasigluons, that are transverse spin-1 bosons in the adjoint representation of the gauge group. The baryonic potential can be set equal to 0 and according to standard formulas in statistical mechanics, one has
\begin{equation}\label{pg}
	\Omega_0=2 \, {\dim }\, adj\, \omega_0(m_g),
\end{equation}
where the quasigluons are \textit{a priori} supposed to have a mass $m_g$, and where
\begin{equation}\label{oo}
	\omega_0(m)=\frac{1}{2\pi^2\beta}\int^\infty_0dk\, k^2 \ln\left(1-{\rm
e}^{-\beta\sqrt{k^2+m^2}}\right)
\end{equation}  
is the grand potential per degree of freedom associated to a bosonic species with mass $m$. Equation (\ref{psb}) leads to    
  \begin{equation}\label{psbg}
	p_{SB}=\frac{\pi^2}{45 \beta^4}{\rm dim}\, adj.
\end{equation}
Let us recall that in the following, the term gluon indifferently denotes the gauge field of Yang-Mills theory and the quasigluons.

The sum $\sum_\nu$ appearing in (\ref{pot0}) now explicitly reads $\sum_{n_g}\sum_{{\cal C}}\sum_{J^{PC}}$, where $n_g$ is the number of gluons involved in the interaction process. As soon as $n_g>2$, the determination of the allowed color channels and of the correct symmetrized gluon states generally becomes a painful task, to which the problem of finding the $T$-matrix in many-body scattering must be added. Intuitively, one can nevertheless expect the dominant scattering processes to be two-gluon ones, and thus only consider $n_g=$ 2 in a first approach. After simplification, the grand potential (\ref{pot0}) eventually reads
\begin{eqnarray}\label{omega2}
&&\Omega^{(2)}=2\,{\dim }\, adj\,\omega_0(m_g)+\sum_{{\cal C}}\sum_{J^{P}}{\rm dim}\, {\cal C}\, (2J+1) \Bigg\lbrace\omega(M_{{\cal C},J^{P}}) 
 \\
&& + \frac{1}{2\pi^2\beta^2}\int^\infty_{2m_g} d\epsilon \, \epsilon^2\,  K_2(\beta\epsilon)\,  {\rm Tr}_{{\cal C},J^{P}}\Big[ ( \delta {\rm Re}{\cal T })'- 2\pi \big( (\delta {\rm Re}{\cal  T}) (\delta {\rm Im}{\cal T})'-(\delta {\rm Im} {\cal  T}) (\delta {\rm Re}{\cal  T})'  \big) \Big]  \Bigg\rbrace , \nonumber
\end{eqnarray} 
where the symbol ``prime" is the derivative respective to the energy and $M_{{\cal C},J^{P}}$ is the mass of the two-gluon bound state with color ${\cal C}$ and quantum numbers $J^{P}$, if it exists. The index $C$ in the $J^{PC}$ channel is dropped since the charge conjugation is always positive for a two-gluon state \cite{Boul}. In the remaining trace, it is understood that the $T$-matrix has been computed in a given two-body channel with color ${\cal C}$ and quantum numbers $J^{P}$, and that the Dirac $\delta$ reads $\delta(\epsilon-2\epsilon(q))$, with the dispersion relation $\epsilon(q)=\sqrt{q^2+m_g^2}$. Note also that, in connection with nuclear many-body approaches, (\ref{omega2}) can be rewritten in terms of a weighted thermal average of scattering phase shifts by means of unitarity of the $S$-matrix. The computation of the two-gluon $T$-matrix is explained in detail in the following section. 

\subsection{Helicity states and the Lippman-Schwinger equation}\label{helicity}
\subsubsection{Two gluon states}

Jacob and Wick's helicity formalism~\cite{jaco} can be applied to describe a two-gluon state, where the gluons are seen as transverse spin-1 particles. Let us generically define $\left|\psi(\vec p,\lambda)\right\rangle=a^{\dagger}_{\lambda}(\vec p\,)\left|0\right\rangle$ the quantum state of a particle with momentum $\vec p$, spin $s$, and helicity $\lambda$. If the particle is transverse, only $\lambda=\pm s$ is allowed, while all the projections from $-s$ to $+s$ are allowed if the particle has a usual spin degree of freedom. Then it can be deduced from~\cite{jaco} that the quantum state
\begin{equation}\label{hstate}
\left|J^P,M;\lambda_1,\lambda_2,\eta\right\rangle=\frac{1}{\sqrt 2} \Big[	 \left|J,M;\lambda_1,\lambda_2\right\rangle+\eta\left|J,M;-\lambda_1,-\lambda_2\right\rangle\Big],
\end{equation}
with $\eta=\pm1$ and
\begin{eqnarray}
\left|J,M;\lambda_1,\lambda_2\right\rangle&=&\left[\frac{2J+1}{4\pi}\right]^{\frac{1}{2}}\int^{2\pi}_0d\phi\int^\pi_0d\theta\, \sin\theta\ {\cal D}^{J*}_{M,\lambda_1-\lambda_2}(\phi,\theta,-\phi)\, R(\phi,\theta,-\phi)\nonumber\\
&&\qquad\qquad\qquad\qquad\qquad\times \, a^{\dagger}_{\lambda_1}(\vec p\,)a^{\dagger}_{\lambda_2}(-\vec p\,)\left|0\right\rangle,
\end{eqnarray}
is a two-particle helicity state in the rest frame of the system which is also an eigenstate of the total spin $\vec J$ and of the parity, \textit{i.e.} $\vec J^{\, 2}=J(J+1)$, $J_z=M$, and $P=\epsilon\, \eta_1\eta_2(-1)^{J-s_1-s_2}$ with $\eta_i$ and $s_i$ the intrinsic parity and spin of particle $i$. Moreover, $J\geq|\lambda_1-\lambda_2|$.  In the above definition, $R(\alpha,\beta,\gamma)$ is the rotation operator of Euler angles $\{\alpha,\beta,\gamma\}$ and ${\cal D}^{J}_{M,\lambda}(\alpha,\beta,\gamma)$ are the Wigner $D$-matrices. The coordinates $\{\theta,\phi\}$ are the polar angles of $\vec p$. When both particles have a spin degree of freedom, the helicity basis, spanned by the helicity states~(\ref{hstate}), is equivalent to a standard $\left|^{2S+1}L_J\right\rangle$ basis up to an orthogonal transformation \cite{gie}. When at least one of the particles is transverse, both basis are no longer equivalent, but the helicity states can still be expressed as particular linear combinations of $\left|^{2S+1}L_J\right\rangle$ states \cite{jaco}. This will be convenient in view of future computations. 

When the two particles are identical ($m_1=m_2=m$, $s_1=s_2=s$), it is relevant to study the action of the permutation operator $P_{12}$. One finds~\cite{jaco}
\begin{equation}\label{symdef}  
\left[1+(-1)^{2s}P_{12}\right]\left|J^P,M;\lambda_1,\lambda_2,\eta\right\rangle=\left|J^P,M;\lambda_1,\lambda_2,\eta\right\rangle+(-1)^{J}\left|J^P,M;\lambda_2,\lambda_1,\eta\right\rangle,
\end{equation}
where the operator $\left[1+(-1)^{2s}P_{12}\right]={\cal S}$ is nothing else than a projector on the symmetric ($s$ integer) or antisymmetric ($s$ half-integer) part of the helicity state. It is readily seen in (\ref{symdef}) that symmetrizing the state will eventually lead to selection rules for $J$ (this is particularly clear if one sets $\lambda_1=\lambda_2$). When extra degrees of freedom are added, it is also of interest to use the antisymmetrizer $\left[1-(-1)^{2s}P_{12}\right]={\cal A}$ as done in Table~\ref{tabs}. 

A general discussion about the two-gluon helicity states can be found in~\cite{heli}, to which we refer the interested reader. For the present work, it is sufficient to recall that four families of helicity states can be found, separated in helicity singlets $\left|S_\pm; J^P\right\rangle$ and doublets $\left|D_\pm; J^P\right\rangle$ following the pioneering work~\cite{barnes}. The corresponding quantum numbers are given in Table~\ref{tabs}, as well as the average value of the squared orbital angular momentum, $\left\langle \vec L^2\right\rangle$, computed with these states.

\begin{table}[t]
\caption{Symmetrized and antisymmetrized two-gluon helicity states, following the notation of~\cite{heli,barnes}, with the corresponding quantum numbers and averaged squared orbital angular momentum.} 
\begin{tabular}{c|ccc}
State & Symmetrized & Antisymmetrized & $\left\langle \vec L^2\right\rangle$\\
\hline
$\left|S_+; J^P\right\rangle$ & (even-$J\geq0$)$^+$ & (odd-$J\geq1$)$^-$ & $J(J+1)+2$ \\
$\left|S_-; J^P\right\rangle$ & (even-$J\geq0$)$^-$ & (odd-$J\geq1$)$^+$ & $J(J+1)+2$ \\
$\left|D_+; J^P\right\rangle$ & (even-$J\geq2$)$^+$ & (odd-$J\geq3$)$^-$ & $J(J+1)-2$ \\
$\left|D_-; J^P\right\rangle$ & (odd-$J\geq3$)$^+$  & (even-$J\geq2$)$^-$ & $J(J+1)-2$ \\
\end{tabular}

\label{tabs}
\end{table}

The averaged orbital angular momentum is an interesting quantity since it helps to globally understand the mass hierarchy of the glueball spectrum~\cite{heli}. Moreover, in a naive nonrelativistic picture, it estimates the strength of the orbital barrier in scattering theory. For obvious numerical reasons, all the possible $J^P$ channels contributing to $\Omega$ can not be included, that is why it is of interest to find the channels that will presumably contribute the most, \textit{i.e.} those with the lowest value of $\left\langle \vec L^2\right\rangle$. First, one has the symmetric states 
\begin{eqnarray}\label{scaps}
    \left|S_+;0^{+}\right\rangle&=&\left[\frac{2}{3}\right]^{1/2}\left|^1 S_0\right\rangle+\left[\frac{1}{3}\right]^{1/2}\left|^5 D_0\right\rangle,\\
    \left|S_-;0^{-}\right\rangle&=&-\left|^3P_0\right\rangle,
\end{eqnarray}    
expressed in a standard $\left|^{2S+1}L_J\right\rangle$ basis, with $\left\langle \vec L^2\right\rangle=2$. In the singlet channel, they correspond to the $0^{++}$ and $0^{-+}$ glueballs respectively, which are indeed found to be among the lightest ones at zero temperature, see \textit{e.g.} the review~\cite{rev}. Then, with $\left\langle \vec L^2\right\rangle=4$, one has the symmetric state
 \begin{eqnarray} \label{tensor} 
    \left|D_+;2^{+}\right\rangle&=&\left[\frac{2}{5}\right]^{1/2}\left|^5S_2\right\rangle
    +\left[\frac{4}{7}\right]^{1/2}\left|^5 D_2\right\rangle+\left[\frac{1}{35}\right]^{1/2}\left|^5 G_2\right\rangle,
\end{eqnarray}
and the antisymmetric states
\begin{eqnarray}
    \left|S_+;1^{-}\right\rangle&=&\left[\frac{2}{3}\right]^{1/2}\left|^1 P_1\right\rangle-\left[\frac{2}{15}\right]^{1/2}\left|^5 P_1\right\rangle+\left[\frac{1}{5}\right]^{1/2}\left|^5 F_1\right\rangle,\\
    \left|S_-;1^{+}\right\rangle&=&\left[\frac{1}{3}\right]^{1/2}\left|^3 S_1\right\rangle-\left[\frac{2}{3}\right]^{1/2}\left|^3 D_1\right\rangle,\\
    \left|D_-;2^{-}\right\rangle&=&-\left[\frac{4}{5}\right]^{1/2}\left|^5 P_2\right\rangle-\left[\frac{1}{5}\right]^{1/2}\left|^5 F_2\right\rangle .
\end{eqnarray}
The above three states do not exist in the singlet channel, but the symmetric $2^+$ state corresponds to the $2^{++}$ glueball in the singlet channel; the $0^{\pm+}$ and $2^{++}$ are indeed the lightest states at zero temperature~\cite{rev}. Only the color-symmetric channels with the lowest value of $\left\langle \vec L^2\right\rangle$ will be kept in the following study, which aims at being a first step toward a description of the Yang-Mills plasma within a $T$-matrix formulation. 

\subsubsection{Lippman-Schwinger equation}

Solving (\ref{ls}) is a crucial technical part of this work since it eventually leads to the on-shell $T$-matrix.  As it will be discussed in Sec.\ref{param}, the potential to be used is known in position space and has firstly to be Fourier-transformed. For a potential with spherical symmetry in configuration space, we use
\begin{equation}\label{Vqq}
V(q,q',\theta_{q,q'}) = 4 \pi \displaystyle\int_0^\infty dr \, r V(r) \displaystyle\frac{\sin(Qr)}{Q}, \quad {\rm where}\quad  Q = \sqrt{q^2 + q'^2 - 2 q q' \cos \theta_{q,q'}}\text{,}
\end{equation}
and where $\theta_{q,q'}$ is the angle between the momenta $\vec q$ and $\vec q\,'$. 

Since two-gluon interactions are considered, the basis states are two-gluon helicity states, given in the above section. As we assume $V$ to be spin independent (see Appendix~\ref{appU1}), only the orbital angular momentum containing of the helicity states has to be taken into account. According to a standard integration, the $L$-wave part of potential (\ref{Vqq}) reads
\begin{equation}
V_L(q,q') = 2\pi \displaystyle\int_{-1}^1 dx P_L(x) V(q,q',x),
\end{equation}
where $P_L$ is the Legendre polynomial of order $L$ and $x = \cos \theta_{q,q'}$. Our choice is to focus on the scalar, pseudoscalar and tensor scattering channels, for which one can compute from (\ref{scaps})-(\ref{tensor}) that
\begin{eqnarray}
V_{0^{+}}(q,q') &=& \displaystyle\frac{2}{3} V_0(q,q') + \displaystyle\frac{1}{3} V_2(q,q')\text{,}\\
V_{0^{-}}(q,q') &=& V_1(q,q')\text{,} \\
V_{2^{+}}(q,q') &=& \displaystyle\frac{2}{5} V_0(q,q') + \displaystyle\frac{4}{7} V_2(q,q') + \displaystyle\frac{1}{35} V_4(q,q') \text{.}
\end{eqnarray}
Note that once $V_{J^P}(q,q')$ is known, that is the potential in a given $J^P$ scattering channel, the off-shell $T$-matrix can be computed from (\ref{ls}) as follows in \cite{cabre06}: 
\begin{equation}\label{tosolve}
{\cal T}(E; q,q') = V_{J^P}(q,q') + \displaystyle\frac{1}{8\pi^3} \displaystyle\int_0^\infty dk\, k^2\, V_{J^P}(q,k) \,G_0(E;k)\, {\cal T}(E;k,q'),
\end{equation}
where the two-gluon propagator reads
\begin{equation}
G_0(E;k) = \displaystyle\frac{m_g^2}{\epsilon(k)} \displaystyle\frac{1}{E^2/4- \epsilon(k)^2 - 2i \, \epsilon(k)\, \Sigma_I}
\end{equation}
with the gluon dispersion relation $\epsilon(k)= \displaystyle\sqrt{k^2+m_g^2}$. Note that the  normalization conventions of the $T$-matrix are not the same as the ones in \cite{cabre06} (see Appendix~\ref{trace}).  The parameter $\Sigma_I$ accounts for the imaginary part of the gluon self-interaction, whereas the real part is reabsorbed in the effective gluon mass. A more complete calculation of the gluon self-energy would require summing the $T$-matrix over the gluon thermal distribution self-consistently in a Brueckner-Hartree-Fock scheme (schematically, $\Pi_g=\int f^g {\cal T} D_g$ with $f^g$ the gluon distribution and $D_g$ the gluon single-particle propagator). We leave such determination for a future work. In the present evaluation, we shall approximate the gluon self-interaction by using an effective, in-medium gluon mass (to be discussed in Sec.~\ref{param}) together with a small imaginary part for numerical purposes (we use $\Sigma_I=0.01$~GeV as in \cite{cabre06}).

Once ${\cal T} (E;q,q')$ is known, the on-shell $T$-matrix is readily obtained
as ${\cal T}(E;q_E,q_E)$, with $q_E=\sqrt{E^2/4-m_g^2}$. The
Haftel-Tabakin algorithm is used to solve (\ref{tosolve}) \cite{haftel}. The
momentum integral is discretized within an appropriate quadrature, thus turning
the integral equation into a matrix equation, namely, $\sum {\cal F}_{ik} {\cal
T}_{kj} = V_{ij}$, where, schematically, ${\cal F}=1-wVG$ (and $w$ denotes the
integration weight). The solution follows trivially by matrix inversion. It can
be shown that the determinant of the transition function $\cal F$ (referred to
as the Fredholm determinant) vanishes at the bound state energies, which provides
a numerical criterion for solving the bound state problem. This strategy has
already been successfully used to compute $T$-matrices in the case of
quark-antiquark scattering \cite{cabre06}.

\section{Thermodynamic observables with SU($N$) and G$_2$}\label{sunT}

\subsection{Pure gauge sector}

\subsubsection{SU($N$) case}

The explicit computation of $\Omega^{(2)}$, given by (\ref{omega2}), obviously requires the knowledge of the on-shell $T$-matrix, that can be derived in particular from (\ref{ls}). In this last equation, $G_0$ is the propagator of two gluons, that has been discussed in Sec.~\ref{YM}. It is only worth saying that $G_0= {\rm O}(1)$ with respect to the number of colors since $m_g$ is assumed to be ${\rm O}(1)$ (see Sec.~\ref{gm}). The color dependence of the $T$-matrix actually comes from the two-gluon interaction potential only. More precisely, the color-dependence of the potential is all included in the factor (\ref{kappa0}), reading in the present case
\begin{equation}\label{kappa0b}
\kappa_{{\cal C};gg}=\frac{C^{{\cal C}}_2-2N}{2N}.
\end{equation}
The subscript $gg$ is used to recall that two-gluon interactions are concerned in the above formula, and $C_2^g = C_2^{adj} = N$ in the SU($N$) case.  

The adjoint representation of SU($N$), to which gluons belong, can be written as the ($N-1$)-component vector $(1,0,\dots ,0,1)$ in a highest weight representation, corresponding to a Young diagram with $1$ column of length $N-1$ and 1 column of length 1. More generally, $(a_1,\dots,a_k,\dots,a_{N-1})$ corresponds to a Young diagram with $a_k$ columns of length $k$. The tensor product of the adjoint representation by itself gives the allowed two-gluon color channels:
\begin{eqnarray}\label{colo}
(1,0,\dots ,0,1)&\otimes &(1,0,\dots,0,1)=\nonumber\\
&&\bullet^S\oplus \, (1,0,\dots,0,1)^A\oplus (2,0,\dots,0,2)^S\qquad\qquad\ \  \hspace{1.5cm} N\geq 2\nonumber \\
& &\oplus \,(1,0,\dots,0,1)^S \oplus (0,1,0,\dots,0,2)^A \oplus (2,0,\dots,1,0)^A \quad N\geq 3\nonumber \\
& &\oplus \,(0,1,0,\dots,0,1,0)^S \hspace{6.4cm} N\geq 4.
\end{eqnarray}
The superscript $S$/$A$ denotes a symmetric/antisymmetric channel. The first/second/third line exists as soon as $N\geq$ 2/3/4. Note that in the special case $N=2$, the above tensor product reduces to  $(2)\otimes(2)=(0)^S\oplus(2)^A\oplus (4)^S$, and one recovers usual spin-coupling rules. The dimensions and color factors of the representations appearing in (\ref{colo}) can be found in Table~\ref{tab1}. 

In the singlet channel, one has $\kappa_{\bullet}=-1$ for any $N$. It is such that ${\cal T}={ \rm O} (1)$ since $V={\rm O}(1)$. Consequently, the properties of glueballs in singlet above the deconfinement temperature are not dependent of $N$, in agreement with \cite{hage}, where it is suggested that this argument is even gauge-group independent. The singlet finally brings a contribution ${\rm O} (1)$ to $\Omega^{(2)}$ since its dimension is 1. 

Using the same arguments as for the singlet, one finds that the adjoint channels also lead to a $T$-matrix that is $N$-independent. They may lead to bound states since the potential is attractive, though less strongly than for the singlet. The symmetric adjoint channel will presumably be the most favorable for the formation of bound states since it demands a completely symmetric spin-space wave function for the two-gluon state in virtue of the Pauli principle, and the most attractive $J^P$ channels are indeed symmetric. Note that this symmetric color channel is actually absent for $N=2$. In the adjoint channel, ${\cal T}= {\rm O}(1)$ since $V= {\rm O} (1)$ but, unlike the singlet, its contribution to $\Omega^{(2)}$ is ${\rm O}(N^2)$ since ${\rm dim} (1,0,\dots,0,1)=N^2-1$. 

The two remaining channels with nonzero potential, namely $(2,0,\dots,0,2)$ (the $\bm{27}$ for SU(3)) and  $(0,1,0,\dots,0,1,0)$ (only when $N>3$), have in common that they are symmetric and that their color factor scales in $1/N$, thus vanishes in the large-$N$ limit. The fact that $V= {\rm O}(1/N)$ in both cases leads to the exact large-$N$ result
\begin{equation}
{\cal T}=V+V\, G_0\, V+{\rm O}(N^{-3}), 
\end{equation}
or
\begin{equation}
{\cal T}=\pm\frac{1}{N}v+\frac{1}{N^2}v\, G_0\, v+{\rm O}(N^{-3}),
\end{equation}
the $\pm$ coming from one channel or another. Because of the weakness of $V$ at large-$N$, one can reasonably suppose that even the attractive channel $(0,1,0,\dots,0,1,0)$ does not lead to the formation of bound states. For the two channels under consideration,
\begin{equation}
\left.  \left({\cal S}S^{-1}\overleftrightarrow{\partial_\epsilon}S \right)\right|_c \propto
\,\partial_\epsilon{\rm Re}\left(\pm\frac{v}{N}+\frac{vG_0v}{N^2}\right)+ {\rm O}(N^{-3}).
\end{equation}
One sees in (\ref{omega2}) that the contributions of both channels have to be summed and, since they are symmetric, the sums on the allowed $J^{P}$ is identical in both cases. This causes the term in $1/N$ to vanish in the trace at large-$N$ limit, the first nontrivial one being in $1/N^2$, leading to an overall contribution to $\Omega^{(2)}$ scaling as $N^2$ because the dimension of both channels scale as $N^4$. 

Although the color singlet is relevant in view of studying glueballs, it does not bring any contribution to the EoS at large-$N$. So, the large-$N$ EoS is dominated by free gluons and scattering processes above threshold in colored channels. The more $N$ is large, the more important is the gap between the confined phase and the deconfined one, whose EoS scales as $N^2$. It is indeed known that the large $N$-case corresponds to a  strongly first-order phase transition ($N=3$ is already weakly first order)~\cite{TcTh0}.  

\subsubsection{G$_2$ case} \label{G2}
Another interesting group under consideration is G$_2$ which is also the best studied gauge group so far beyond SU($N$). The main features of this group are summarized in what follows. 
\par The adjoint representation of G$_2$ has dimension 14, and reads $(0,1)$ in a highest weight representation. The two-gluon channels are then  given by
\begin{equation}
(0,1)\otimes (0,1)=\bullet^S +(0,1)^A +(0,2)^S+(2,0)^S+(3,0)^A
\end{equation}
or, in terms of the dimensions, $14\otimes 14=1+14+77'+27+77$. Using the same normalisation than in the SU($N$) case, the color factors respectively read \cite{lipi,buiss11} $\kappa_{{\rm C};gg}=-1$, $-1/2$, 1/4, $-5/12$, and 0. The color factors in the singlet and adjoint channels are equal to those of SU($N$), so the glueball properties are unchanged in the singlet and antisymmetric adjoint channels. The symmetric $(2,0)^S$ channel is almost as attractive as the adjoint one: It may lead to bound states.

\subsubsection{Scaling relations for SU($N$) and G$_2$} \label{relations}
Some interesting relations about the scaling of the EoS can be deduced thanks to the $T$-matrix. 
Let us write the on-shell $T$-matrix as ${\cal T}= \sum_k a_k\,\kappa_{{\cal C};gg}^k $ where all $a_k$ do not depend on the color but rather on the other quantum numbers involved. The color dependence of the thermodynamic observables is then given by the quantities $\sum_{{\cal C};A/S} \text{dim} \, {\cal C}\,\kappa_{{\cal C};gg}^k$. Using the results of Appendix~\ref{sun} and Sec.~\ref{G2}, one can check that, for SU($N$) and G$_2$, 

\begin{eqnarray}
\label{sumdimk}
\sum_{{\cal C;S}} \text{dim} \, {\cal C}_{gg} \kappa_{{\cal C};gg} &=& \displaystyle\frac{1}{2} \, \text{dim} \,{adj} \text{,} \\
\sum_{{\cal C;S}} \text{dim} \, {\cal C}_{gg} \kappa_{{\cal C};gg}^2 &=& \displaystyle\frac{3}{4}\,\text{dim} \,{adj} \text{,} \\
\sum_{{\cal C;S}} \text{dim} \, {\cal C}_{gg} \kappa_{{\cal C};gg}^3 &=& -\displaystyle\frac{1}{8}\,  \text{dim}\, {adj}\text{,}\\
\sum_{{\cal C;A}} \text{dim} \, {\cal C}_{gg} \kappa_{{\cal C};gg}^k &=& \left(- \displaystyle\frac{1}{2}\right)^k \, \text{dim} \,{adj} \text{.} 
\end{eqnarray}
\par For SU($N$) at large-$N$, the previous relations can be written
\begin{eqnarray} \label{KcR}
\sum_{{\cal C;S}} \text{dim} \, {\cal C}_{gg} \kappa_{{\cal C};gg}^k&=& N^2 \left[ \left(-\frac{1}{2} \right)^k+ \delta_{k,1} + \frac{1}{2}\delta_{k,2} \right] +{\rm O}(1) \text{,} \\
\sum_{{\cal C;A}} \text{dim} \, {\cal C}_{gg} \kappa_{{\cal C};gg}^k &=& N^2\left(-\frac{1}{2} \right)^k +{\rm O}(1)\text{.}
\end{eqnarray}
Again that means that the expected scaling like $N^2$ (actually like dim $adj$) of the EoS is found using the present approach. This can be viewed as a confirmation of the relevance of the chosen color scaling (\ref{V00}).

\subsection{Quarks and antiquarks in the 't~Hooft limit}

Even if the rest of this study will be concerned with a genuine Yang-Mills plasma, it is worth making some comments about the possible inclusion of matter (quarks and antiquarks) in the model. Informations about the color channels appearing in interactions involving at least one (anti)quark are given in Tables \ref{tab3} and \ref{tab4}; remark that quarks (antiquarks) have been considered to be in the fundamental (conjugate) representation of SU($N$), as it is the case in 't~Hooft large-$N$ limit \cite{hoof}. Other interesting large-$N$ limits have been proposed, in which quarks belong to the two-index antisymmetric representation of SU($N$) for example \cite{qcdas}, but they will not be studied here. 

First of all, the quark-quark and antiquark-antiquark color factors are of order $1/N$: The Born approximation for the $T$-matrix becomes exact at large-$N$ (at any temperature) and, since the dimension of any of the corresponding representations scales as $N^2$, the quark-quark and antiquark-antiquark interactions bring a term scaling as $N$ to the grand potential. More precisely, this term scales as $N_f \, N$ at the Born approximation once the trace over the different flavors is performed. It is shown in Sec.~\ref{Born} that only the interaction of two identical species can contribute to $\Omega$ in this limit.  The number of quark flavors remains finite in the 't~Hooft limit, which is the case under study here.

The quark-antiquark interactions lead to a $T$-matrix which is ${\rm O}(1)$ when the pair is in the singlet, or ${\rm O}(1/N^2)$ when the pair is in the adjoint representation. In both cases however, the contribution to the grand potential scales as $N_f(N_f + 1)/2$ at large-$N$. So the quark-antiquark contributions to the thermodynamic observables is negligible with respect to the quark-quark and antiquark-antiquark ones in the 't~Hooft limit. 

Finally, using similar arguments, one can show that the contribution of the quark-gluon and antiquark-gluon interactions to the grand potential scale as $N_f \, N$ at large-$N$. One concludes that, in the 't~Hooft large-$N$ limit, the grand potential is dominated by the gluonic contributions only, scaling as $N^2$.

\section{Parameters of the model}\label{param}

\subsection{Potential and gluon mass}\label{gm}

Two ingredients are now missing to start numerical computations: The interaction
potential between two gluons and the gluon mass. The procedure followed to fix
the potential is similar to the one followed in~\cite{cabre06} in the case of
heavy quark-antiquark bound states. The first step is to take some input from
lattice QCD, from which accurate computations of the static free energy of a
quark-antiquark pair bound in a color singlet, $F_1(r,T)$, are available. In
particular, computations in quenched SU(3) lattice QCD can be found in
\cite{kacz3}; they are especially relevant for our purpose since we focus on the
pure Yang-Mills plasma. There is still debate on the proper potential term to
use in phenomenological approaches, namely $F_1$ or the internal energy
$U_1=F_1-T\partial_T F_1$. An entropic contribution is subtracted from the free
energy in $U_1$, causing the internal energy to be more attractive than the free
energy, eventually leading to larger dissociation temperatures for bound states
in the deconfined medium. Spectral function analysis of heavy quarkonia from
lattice QCD simulations of Euclidean correlation functions typically suggest
that the $\eta_c$ and $J/\psi$ states may survive up to about 2$T_c$. Such
values of the dissociation temperature can be accommodated if the singlet
internal energy is used in potential model calculations.
\cite{Asakawa:2003re,Alberico:2005xw,Wong:2006bx,Riek:2010fk}.
As in~\cite{cabre06},  the internal energy is used as potential term. The explicit expression of the internal energy $U_1$ used in this work can be found in Appendix~\ref{appU1}.

The assumed color scaling (\ref{V0}) allows to derive the two-gluon potential from the lattice quark-antiquark one. Indeed, given $U_1(r, T)$ computed in quenched SU($N_{lat}$) lattice data, the color factor of the singlet quark-antiquark pair reads
\begin{equation}
\kappa_{q\bar q}=-\frac{N^2_{lat}-1}{2N_{lat}^2}.
\end{equation}
According to (\ref{V0}), the potential (in position space) between two quasigluons in the color channel ${\cal C}$ is then given by
\begin{equation}\label{Vg}
V(r,T)=\frac{\kappa_{{\cal C};gg}}{\kappa_{q \bar q}} \left[U_1(r,T)-U_1(\infty , T)\right],
\end{equation}
where the long-distance limit of the potential has be normalized to zero in order to ensure the convergence of the scattering equation and to perform the Fourier transform. This is actually a standard procedure in finite-temperature calculations. 

According to the suggestion made in ~\cite{mocsy05}, the nonzero value of $U_1(\infty,T)$ should eventually be responsible of an effective in-medium contribution to the gluon mass. The intuitive argument is that, when both gluons are infinitely separated, they no longer interact. Therefore, the remaining potential energy should be seen as a manifestation of self-energy effects induced by the surrounding medium. These effects are encoded in the model as a mass shift to the ``bare" quasigluon mass, whose value has still to be fixed. Since $U_1(\infty,T) = 2\,m_q(T)$, the adaptation to the gluon must be done by extracting the correct color-dependence. From HTL computations \cite{HTL}, the self-energy color dependence is given by $C_2^ q/C_2^ {adj}$ at the first order when it is added in the propagator as a mass term ($m^ 2$), that means here that
\begin{equation}
\displaystyle\frac{U_1(\infty, T)}{2} = m_q(T) = \displaystyle\sqrt{\displaystyle\frac{C_2^q}{C_2^{adj}}} \Delta(T).
\end{equation}
So, $m_q(T) = 2\Delta(T)/3$ in the SU(3) case. 

\par In the same way as done for the two-body color scaling, $\Delta(T)$ is considered as universal and the gluon thermal mass reads
\begin{equation} \label{scalingM}
\delta(T) = \displaystyle\sqrt{\displaystyle\frac{C_2^g}{C_2^{adj}}} \Delta(T) = \Delta(T),
\end{equation}
since $C_2^g = C_2^{adj}$. So, $\delta(T)$ is gauge-group independent. The effective in-medium gluon mass is finally given in our approach as 

\begin{equation}\label{mg}
m_g(T)^2=m_0^2+\delta(T)^2.
\end{equation}
where the value $m_0$ has still to be fitted (see the following section). All the contributions are quadratically added  as it is the case when one is dealing with bosonic propagators. The gluon mass is thus gauge-group independent. 
\par It is obvious that the problem of the gluon mass is far more complicated than the simple prescription (\ref{mg}), that has to be seen as valid in a first approximation only. A more refined gluon mass should probably be momentum-dependent. There is indeed an increasing amount of evidences favoring the existence of a dynamically generated gluon mass due to nonperturbative effects, at least at zero temperature. Such a dynamically generated gluon mass $m_g(p)$, with $m_g(\infty)=0$ and $m_g(0)$ finite, is favored by some lattice results in Landau gauge, see \textit{e.g.} \cite{gluml1,gluml2}. Also nonperturbative field-theoretical calculations, using for example the pinch technique, find a nonzero dynamically generated gluon mass in $3+1$ YM theory \cite{glumass,glumass2}. It is also worth quoting the recent Coulomb gauge study \cite{mgfinit}, which is a first step in view of understanding the behavior of $m_g(p,T)$ at a nonperturbative level. From a different perspective, non-perturbative contributions to the gluon potential and mass are analyzed at finite
temperature in connection with the gluon condensates in \cite{Megias:2007pq,Megias:2009mp}.
Such improvements of the gluon mass are left for future works. 

The above discussion gives a more precise meaning to the term ``quasigluon" used in this paper: It denotes transverse particles in the adjoint representation of SU($N$) that gain an effective mass $m_g(T)$ given by (\ref{mg}) and interact through the potential (\ref{Vg}).

\subsection{Zero temperature results with SU(3)} \label{zeroT}

Before performing finite-temperature computations, it is worth checking whether the values retained for the various parameters of our model may give relevant results at zero temperature or not. In particular, is the present $T$-matrix formalism able to reproduce at least qualitatively the features of the low-lying glueball spectrum computed in pure gauge SU(3) lattice QCD  at zero temperature ? \cite{glulat} 

It is known that, at zero temperature and in quenched SU(3) lattice QCD, the potential between a static quark-antiquark pair is compatible with the funnel form \cite{bali}
\begin{equation}
V_f(r) = \sigma r -\frac{4}{3}\frac{\alpha}{r}.
\end{equation}
In order to stay coherent with the potential  above $T_c$, $\alpha=0.141$ (see Appendix \ref{appU1}) and $\sigma = 0.176$~GeV$^2$ (a standard value for the string tension) are used. The Fourier transform of $V_f(r)$ is not defined: This flaw can be cured by making it saturate at some value $V_{sb}$, interpreted as a string-breaking value, that is the energy above which a light quark-antiquark pair can be created from the vacuum and break the QCD string. This scale is then subtracted and the potential effectively taken into account is $V_f(r)-V_{sb}$, while $V_{sb}$ is interpreted as an effective quark mass using the same arguments as those detailed in Sec.~\ref{gm}. According to the color scaling (\ref{kappa0}), the potential that should be used to describe the interactions between two gluons at zero temperature is  
\begin{equation}
V_0(r) = \displaystyle\frac{9}{4} V_f(r) - V_{sb}^g
\end{equation}
when the gauge group is SU(3). 
In this case, the string breaking scale should rather be interpreted as the energy scale necessary to form two gluelumps, a gluelump being a gluon bound in the color field of a static adjoint source. It is known indeed that adjoint string breaking may be observed, and occurs at twice the lightest gluelump mass ($\sim 2$~GeV) \cite{defor}. So, $V_{sb}^g = 2$~GeV is used here, a value in agreement with lattice data showing that the mass of the lightest gluelump is given by $0.85(17)$~GeV \cite{pineda}. 

The only free parameter left to compute the ${T}$-matrix is the bare gluon mass, $m_0$.  
Keeping the same structure as in Eq.~ (\ref{mg}) we have 
\begin{equation}
m_g(0)^2 = m_0^2 + \left(\displaystyle\frac{V_{sb}^g}{2}\right)^ 2,
\end{equation}
where again we have traded the subtracted potential at infinite separation distance (string breaking energy in this case) into a self-energy-like contribution to the quasi-particle gluon mass.
We fix $m_0 = 0.7$~GeV, which is a typical value for the zero-momentum limit of the gluon propagator at zero temperature. Advancing results, such a value will ensure both a correct agreement with the zero temperature lattice glueball spectrum (see Table \ref{tabglueb}), and an excellent agreement with the pure gauge EoS computed on the lattice (see Sec. \ref{EoS}).

\begin{table}[ht]
\caption{Masses (in GeV) of the lowest-lying glueball states at zero temperature with the gauge group SU(3). Our results (third column), are compared to the lattice data of \cite{glulat} (second column) and to the Coulomb gauge QCD (CGQCD) study \cite{cg0} (last column). } 
\begin{tabular}{c|rrr}
State & Lattice \cite{glulat} & $T$-matrix & CGQCD \cite{cg0} \\
\hline
$0^{++}$ & 1.73 (5)(8) &  2.17 & 1.98\\
$0^{-+}$ & 2.59 (4)(13) & 2.39  & 2.22 \\
$2^{++}$ & 2.40 (2.5)(12) & 2.34  & 2.42 \\
\end{tabular}

\label{tabglueb}
\end{table}

The results are given in Table~\ref{tabglueb} for $J^P$ channels discussed in Sec~\ref{helicity}. At least our model is able to reproduce the mass hierarchy of the lightest glueballs as well as the typical mass scale of 2~GeV for those states. The accuracy of the model can be compared to Coulomb gauge QCD \cite{cg0}, sharing formally many similarities with our $T$-matrix approach: The results of this last reference are also given in Table~\ref{tabglueb}. The agreement between lattice QCD and CGQCD is better, but it is worth saying that the parameters used in CGQCD have been chosen to reach an optimal agreement with the zero temperature lattice data, while here the values are mostly designed to give good results above $T_c$. So our model results are satisfactory in that sense. What makes us to find such a high mass for the scalar glueball is the quite small value $\alpha=0.141$ that has been taken (in order to fit static potentials above $T_c$), while values as high as $\alpha=0.4$ have sometimes to be used to reach a good agreement between lattice data and effective approaches, see \textit{e.g.} \cite{gluheli}. The scalar glueball being dominantly a $S$-wave state, it is particularly sensitive to the strength of the Coulomb term and to the running of $\alpha$ with the temperature, that is neglected here. 

Finally, the extension of the above calculations to any gauge group is straightforward in our approach: The interested reader will find a discussion of such a generalization in \cite{buiss11}, where it is shown that the lowest-lying glueball masses is gauge-group independent within a constituent framework. In particular, the lowest-lying glueball masses are found independent of $N$ in \cite{buiss11}, in agreement with what is observed on the lattice \cite{gluLNC}. That is why the $T$-matrix masses given in Table~\ref{tabglueb} are considered as valid for any gauge group too. 

\subsection{Relevance of the parameters for other gauge groups}

At this stage, it is important to summarize the various parameters introduced and their possible dependence -- or not -- on the gauge group. Comparison with existing results when it is possible can also shed light on that issue.

First of all, the basic ingredient underlying the present study is the static quark-antiquark potential computed in finite-temperature quenched lattice QCD. The assumed one-gluon-exchange nature of the two-particle interactions leads to the universality of the momentum-dependent part of the potential, and to a well-defined prescription for its gauge-group dependence. Similarly, the gluon thermal mass has a peculiar color scaling originating in its interpretation as a self-energy term.

More freedom is apparently left for the numerical parameters at our disposal. Let us comment them briefly. First, the propagator imaginary part $\Sigma_I$ has been introduced for computational convenience. Hence it can be kept constant when changing the gauge group. Second, it can be checked that, by dimensional analysis, our results can all be expressed in terms of the ratios $T/T_c$, $T_c/\sqrt\sigma$ and $m_0/\sqrt\sigma$.

According to glueball gas models with a Hagedorn spectrum describing the high-lying glueball states, the critical temperature is given by \cite{meyer,hage}

\begin{equation}
\label{Tcsig}
\frac{T_c}{\sqrt\sigma}=\sqrt{\frac{3}{2\pi}}=0.69.
\end{equation}

This value is due to the Hagedorn spectrum, not defined above a certain temperature. This temperature is here interpreted as the deconfinement one. It is worth saying that it leads to an EoS in very good agreement with lattice results \cite{meyer,hage,pane2} below $T_c$. In this picture, the ratio $T_c/\sqrt\sigma$ is gauge-group independent: This is only valid in a first approximation since, for example, there are lattice evidences showing that $T_c/\sqrt\sigma$ is only constant up to $1/N^2$ corrections \cite{TcTh}. Nevertheless, such deviation are beyond the scope of this exploratory work. Note that according to \cite{Braun}, the critical temperature is found to be pretty close to 300~MeV up to fluctuation of about 10\% for the gauge groups SU($N$), Sp(2), and E$_7$. So, we fixed $T_c = 300$~MeV in our calculations. This value is in good agreement with (\ref{Tcsig}) for the value $\sigma = 0.176$~GeV$^2$ chosen for $T=0$ calculations (see Sec.~\ref{zeroT}).

Concerning the ratio $m_0/\sqrt{\sigma}$, it is worth mentioning the work \cite{maas}, in which it is shown that the nonperturbative gluon propagator at zero temperature (thus $m_0$ in particular) shows no significant quantitative differences when expressed in units of the string tension for the groups SU($N$) and G$_2$. It is thus tempting to say that the ratio $m_0/\sqrt\sigma$ may be gauge-group independent also: This is assumed in the rest of this paper. The value $m_0/\sqrt\sigma=1.67$ obtained from the zero-temperature glueball spectrum is retained.

Let us finally mention that, at zero temperature, the string breaking scale is found to be two times the gluelump mass for SU($N$) and G$_2$ \cite{lipi}. This means that the extension of the above zero-temperature calculations to any gauge group is straightforward in our approach: The $T$-matrix masses given in Table~\ref{tabglueb} can be considered as valid for any gauge group once divided by $\sqrt\sigma$.

\section{Existence of glueballs above $T_c$}\label{BSsec}

\subsection{Singlet states}
Now that the parameters of the model have been all fixed, $T$-matrix
calculations above $T_c$ can be performed. Technical details will not be given
here since the method is identical to the one used in \cite{cabre06}, which
involves the Haftel-Tabakin algorithm to solve the $T$-matrix Lippmann-Schwinger
equation \cite{haftel}.  The bound and resonant states appear as poles in the
on-shell $T$-matrix, or more precisely as zeros of $\det {\cal F}$. The
corresponding masses are given in Table~\ref{tab:bsatc} in the
different considered color-singlet channels. Since in this case, $\kappa_{{\rm
C};\bullet}=-1$ for all gauge groups and since the gluon mass is independent of
$N$, the masses of the color-singlet are the same for SU($N$) for all $N$, and for 
$G_2$. 

Only a few papers devoted to the existence of glueballs on the lattice are currently known \cite{suga3}, and the interpretation of their results depends mostly on the way the glueball correlators are fitted: Either a single narrow pole, or a Breit-Wigner shape. Let us focus on the narrow pole fit, which identifies bound states in a way similar to ours. The main observation to be made from \cite{suga3} is that the glueball masses decrease above $T_c$ with increasing temperature, with a mass near $T_c$ that is similar to the zero temperature one. This nontrivial behavior is well-checked within our approach. Two competing effects are responsible for the temperature evolution of the spectrum: reduction of the binding energy and downward shift of the threshold energy. Overall, the singlet scalar bound state experiences a mild shift to
lower energies and dissociates at $T_{\textrm{dis}} \approx 1.3~T_c$. This is the value from which
$\det {\cal F}$ does not vanish anymore. Nevertheless, considerable strength remains at
threshold up to about $1.5~T_c$. This is in qualitative agreement with the spectral function
analysis of Euclidean correlators by the CLQCD Collaboration \cite{CLQCD}. 

\begin{figure}[ht]
\begin{center}
\includegraphics*[width=0.6\textwidth]{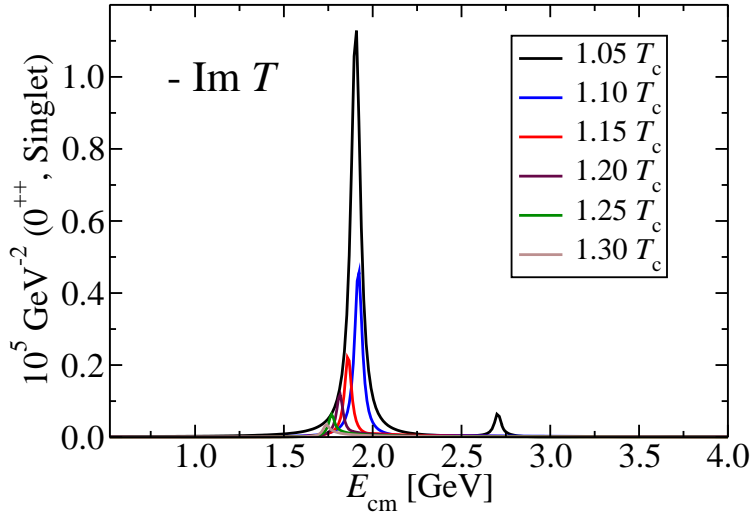}
\caption{
$\textrm{Im} T$ for $gg$
scattering in the
$0^{++}$ singlet channel for various $T$ with $T_c = 0.3$~GeV. 
}
\label{fig3bs}
\end{center}
\end{figure}

\begin{table}
\caption{Masses (in units of $\sqrt{\sigma}$) of lowest-lying glueballs above $T_c$ ($T_c = 0.3$~GeV). A line mark the temperature at which a bound state is not detected anymore.}
\begin{center}
\begin{tabular}{c|c|c|c|c|c|c|c|c|c|c}

\multicolumn{2}{c|}{Channel} & \multicolumn{3}{|c|}{Singlet} & \multicolumn{3}{|c|}{Adjoint$^S$ } & \multicolumn{3}{|c}{(2,0)$^S$ } \\ \hline
\multicolumn{2}{c|}{Group} & \multicolumn{3}{|c|}{All} & \multicolumn{3}{|c|}{SU($N \geq 3$)} & \multicolumn{3}{|c}{$G_2$} \\ \hline
$T/T_c$ & 2\,$m_g$  & $0^{++}$ & $0^{-+} $  & $2^{++}$ & $0^{++}$ & $0^{-+} $  & $2^{++}$ & $0^{++}$ & $0^{-+} $  & $2^{++}$ \\ \hline

1.05 & 6.50 & 4.52 & 5.43 & 5.43 & 6.00 & 6.45 & 6.31  & 6.14 & - & 6.38 \\ 
 &  & 6.48\footnote{Radial excitation below the threshold} & &  &  &  &   &  &  &  \\ 
1.10 & 5.24 & 4.57 & 5.21 & 5.00 & 5.14 & - & - & 5.21 &  & - \\ 
1.15 & 4.71 & 4.43 & -  & 4.67 & - &  &  & - &  &  \\ 
1.20 & 4.43 & 4.33 &  \, & - &  \, &  \, & \, &  \,  &  \,  &  \,  \\ 
1.25 & 4.26 & 4.24 &  \, &  \,&   \, &  \, &  \, &  \,  &  \, &  \, \\
1.30 & 4.14 & - & &  \, &  \, &  \, &  \, &  \, &   \, &  \\
1.35 & - & & &  \, &  \, &  \, &  \, &  \, &   \, & \\

\end{tabular}
\end{center}
\label{tab:bsatc}

\end{table}
The evolution of the imaginary part of the on-shell $T$-matrix in the singlet
scalar channel versus the temperature is displayed in Fig. \ref{fig3bs}: This
gives an overall picture of the glueball progressive dissolution in the medium.
The peak in the imaginary part, depicting a bound state, becomes broader and
broader  before melting into the continuum (and thus $\det {\cal F}$ does not
vanish anymore below threshold) as the temperature is increased. Still, for $T
> T_{\textrm{dis}}$ and above the threshold energy one finds sizable strength from the
bound state relic, the $T-$matrix exhibiting a resonant behavior well beyond the
Born approximation.

Concerning the pseudoscalar channel, singlet bound states are found up to 1.10 $T_c$. Note that states in the pseudoscalar
channels, which in our approach correspond to pure $P$-wave scattering, are
just mildly bound due to the centrifugal barrier. The tensor states, having an $S$-wave component, lie between the scalar and pseudoscalar channels, regarding
binding and dissociation temperatures. 

\subsection{Colored states}

\begin{figure}
\begin{center}
\includegraphics*[width=1.0\textwidth]{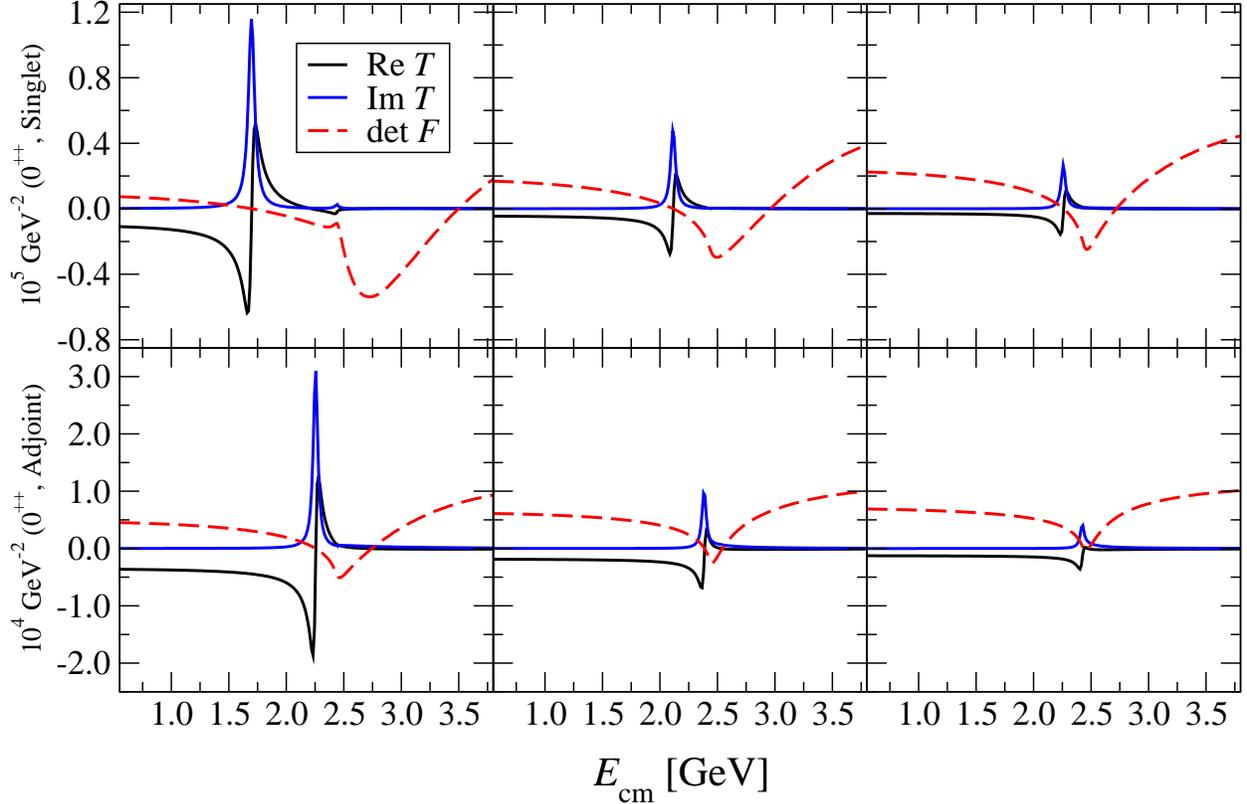}
\caption{ $T$-matrix for $gg$ scattering in the scalar singlet and symmetric adjoint channels for SU($N \geq 3$). From left to right the temperatures are
(1.05;1.10;1.15) $T_c$, with $T_c = 0.3$~GeV. 
}
\label{fig2bs}
\end{center}
\end{figure}
Bound states in the symmetric adjoint channel of SU($N \geq 3$) are also observed (see Table~\ref{tab:bsatc}), although they are less bound since $\kappa_{{\rm C};\bullet}=-1/2$. The scalar channel disappears above 1.10 $T_c$, whereas in the pseudoscalar and tensor channels, bound states are lying right below the threshold energy at the lowest considered temperature (\textit{i.e.} $1.05~T_c$). The differences between singlet and adjoint channels have to be attributed to the strength of the potential, which is two times smaller in the adjoint channel than in the color singlet. 

The evolution of the $T$-matrix in the singlet and symmetric adjoint scalar channel versus the temperature is displayed in Fig.~\ref{fig2bs}. One clearly sees the disappearance of this bound state at 1.15 $T_c$ while the singlet state is still well bound at this temperature.

There are in general other colored channels than the adjoint one. For SU($N$) gauge groups in particular, the only one that could \textit{a priori} lead to bound states is the $(0,1,0,\dots,0,1,0)^S$ channel, which is weakly attractive and exists only for $N>3$. It has been checked that even the scalar state (the most attractive channel) is unbound at $N=4$. Hence this color channel does not admit bound states.
In the case of $G_2$, that is the other group considered in this study, the $(2,0)^S$ channel leads to bound states with the same melting temperatures as in the adjoint channel, up to our current precision of 0.05 $T_c$ (see Table~\ref{tab:bsatc}).  \\

\section{Equation of state}\label{EoS}

The pure gauge EoS can now be computed without introducing extra parameters: The
two-body potential and the thermal mass contribution to the gluon mass have been
fitted on lattice data by using respectively the scaling (\ref{Vg}) and
(\ref{scalingM}), and they ensure a correct agreement with zero temperature
results. A crucial point to establish the EoS thanks to (\ref{omega2}) is to
correctly express the summation on the different channels. For 2-gluon
interactions, the channels are cued by the $ J^{P}$ number and the color number.
The summation on the $J^{P}$ channels is formally infinite but in this work,
only the $0^{++}$, $0^ {-+}$ and $2^{++}$ channels are taken into account. This
restriction is supported by the following argument. These three channels are the
most attractive ones and generate the lightest glueballs. The 
lighter the mass, the more important 
the thermodynamic contribution is in the bound state sector.
Thus, the bound state thermodynamic contribution coming from other
$J^{P}$ should be negligible in comparison of these three channels. In order to be
coherent, this $J^{P}$ restriction is implemented in the scattering sector. It
creates also a restriction in the allowed color channels. Since the $0^{++}$,
$0^ {-+}$ and $2^{++}$ are symmetric $J^{P}$ channels, the color channels must
be symmetric too in order to respect the Pauli's principle. 
\begin{figure}[ht]
\begin{center}
\includegraphics*[width=0.8\textwidth]{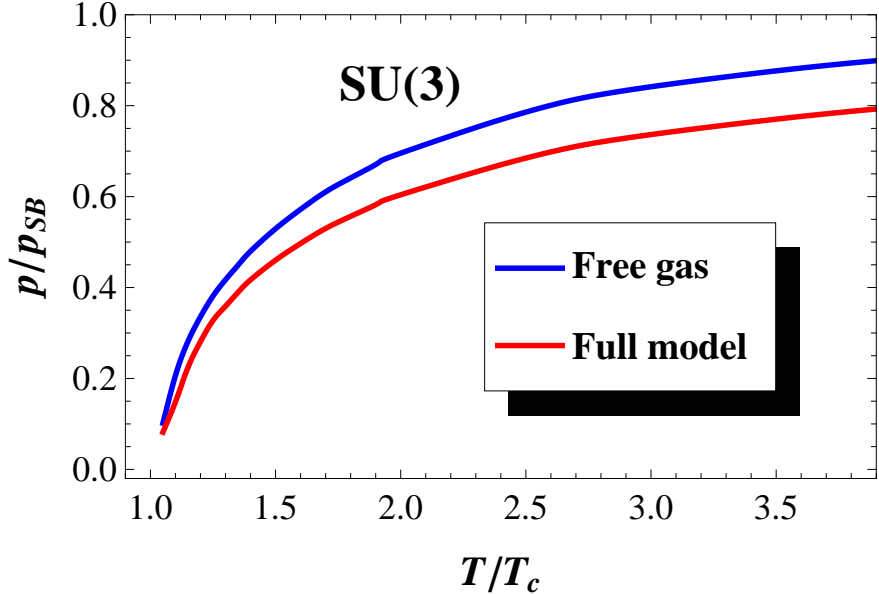}
\caption{Normalized pressure versus temperature in units of $T_c$ (with $T_c$ = 0.3~GeV), computed for the gauge group SU(3) in the free gluon gas case and in the full approach.
}
\label{fig3}
\end{center}
\end{figure}

\par In Fig.~\ref{fig3}, the normalized pressure $P/P_{SB}$ computed in the free gluon gas
case with the thermal mass (\ref{mg}) is compared with the normalized pressure
obtained by our approach in the SU(3) case. At low temperature ($T \leq $ 1.3
$T_c$), the bound state and the scattering parts both give thermodynamic
contributions that modify the free gas pressure. For $T > 1.3 T_c$, only the
scattering part keeps to contribute. As it can be observed in Fig.~\ref{fig3},
the main global effect of the interaction is to decrease the pressure.  If each
contribution is analyzed, it is seen that the bound state formation increases
the  pressure because bound states are simply added as new species that does not
interact with the other particles inside the  plasma. Concerning the two-gluon
scattering part, the sign of the pressure contribution can not be analytically
predicted at each temperature. Only at the Born approximation, one can observe
that attractive (repulsive) channels contribute to increase (decrease) the
pressure. Indeed, in momentum space representation (see Appendix~\ref{trace}),
(\ref{pot_s3}) becomes here
\begin{equation} \label{Born0}
\Omega_s = \displaystyle\frac{1}{64\pi^5 \beta} \displaystyle\sum_{J^P} (2J+1) \displaystyle\sum_{{\cal C},  g} \text{dim} \, {\cal C} \,\kappa_{{\cal C}, gg} \displaystyle\int_{2m_g}^{\infty} d\epsilon \, \epsilon^3 \displaystyle\sqrt{\frac{\epsilon^2}{4} - m_g^2} K_1(\beta \epsilon) \, v_{J^{P}}
\end{equation}
where $K_1(x)$ is the modified Bessel function of the first kind. In the attractive (repulsive) channels, the sign of the potential is negative (positive). Since $\Omega_s$ is the scattering contribution to the grand potential, it can be deduced that attractive (repulsive) channels increase (decrease) the pressure. In the present SU(3) case, the only repulsive channel is the $(2,2)^S$. That means that the decreasing of the pressure in our approach compared with the free gas pressure is only driven by the $(2,2)^S$ channel. 

\par It is also worth wondering whether some constraints arise or not from the
high-temperature limit of our framework concerning the behavior of the two-body
interactions. In this limit, the Born approximation should be relevant. Using
(\ref{sumdimk}) and (\ref{Born0}), one can write 
\begin{eqnarray}
\Omega_s &\sim &\displaystyle\frac{1}{64\pi^5 \beta}  \displaystyle\frac{\text{dim}\,adj}{2}\displaystyle\int_{2m_g}^{\infty} d\epsilon \, \epsilon^3 \displaystyle\sqrt{\frac{\epsilon^2}{4} - m_g^2} K_1(\beta \epsilon) \, v_{0} .
\end{eqnarray}
Only the scalar channel has been taken into account for the sake of clarity, but the following argument can be extended to any spin. According to Hard-Thermal-Loop results, it is relevant to assume a Yukawa form for the potential $v_0$ at high temperature \cite{HTL}. Then,
\begin{eqnarray}
\Omega_s &\sim &\displaystyle\frac{1}{64\pi^5 \beta^4}  \displaystyle\frac{\text{dim}\,adj}{2} \displaystyle\int_{2 \beta m_g}^{\infty} dx \, x^3  K_1(x) \, \frac{\sqrt{\frac{x^2}{4} -\beta^2 m_g^2}}{\frac{x^2}{4} -\beta^2 m_g^2-\beta^2 M^2},
\end{eqnarray}
where $M$ is the screening mass of the theory, \textit{a priori} temperature-dependent. Still in  HTL theory, it is found that, because of the running of the strong coupling constant, 
\begin{equation}\label{limmu}
{\rm lim}_{\beta\rightarrow 0} \beta m_g ={\rm lim}_{\beta\rightarrow 0} \beta M=0. 
\end{equation}
More precisely, the quark and gluon thermal masses are found to behave as $\alpha_s(T) \, T$. 
Consequently, at high enough temperatures, it is found that 
\begin{eqnarray}
\Omega_s &\sim &\ \displaystyle\frac{\text{dim}\,adj}{32 \pi^ 5\beta^4},
\end{eqnarray}
\textit{i.e.} a scattering contribution that has the same behavior with respect to the temperature as the free part, ensuring a well-defined large-temperature limit. 

\begin{figure}[ht]
\begin{center}
\includegraphics*[width=0.8\textwidth]{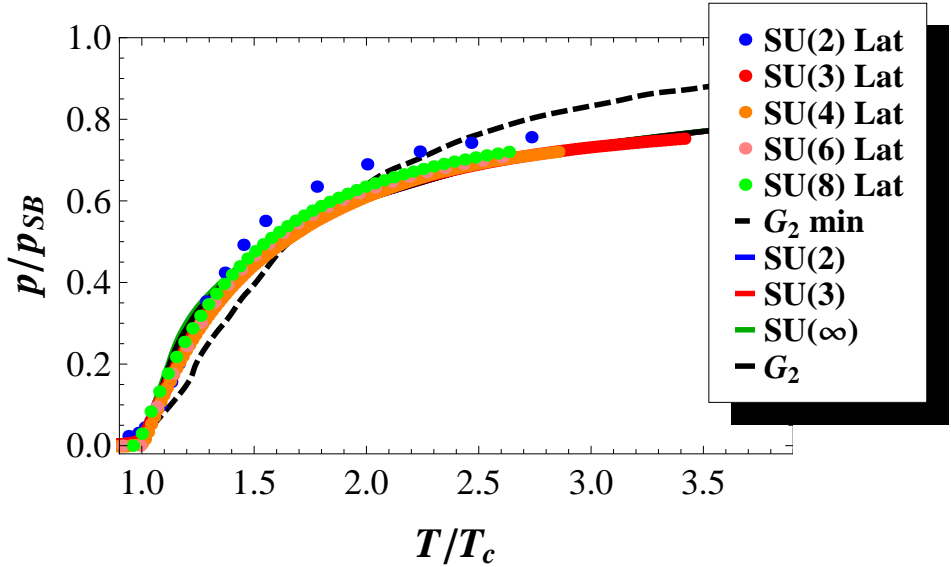}
\caption{Normalized pressure versus temperature in units of $T_c$ (with $T_c$ = 0.3~GeV), computed for the gauge groups SU(2,3,$\infty$) and $G_2$ (solid lines). Note that all the curves are nearly indistinguishable. Our results are compared to the lattice data of \cite{su2lat} for SU(2) (dots) and \cite{panero} for SU(3,4,6,8) (dots), and of the minimal $G_2$ model of \cite{dumi} for $G_2$ (dashed line). Note that all Lattice data have been normalized to the lattice Stefan-Boltzmann pressure \cite{su2lat,panero}.  
}
\label{fig4}
\end{center}
\end{figure}

\begin{figure}[h]
\begin{center}
\includegraphics*[width=0.8\textwidth]{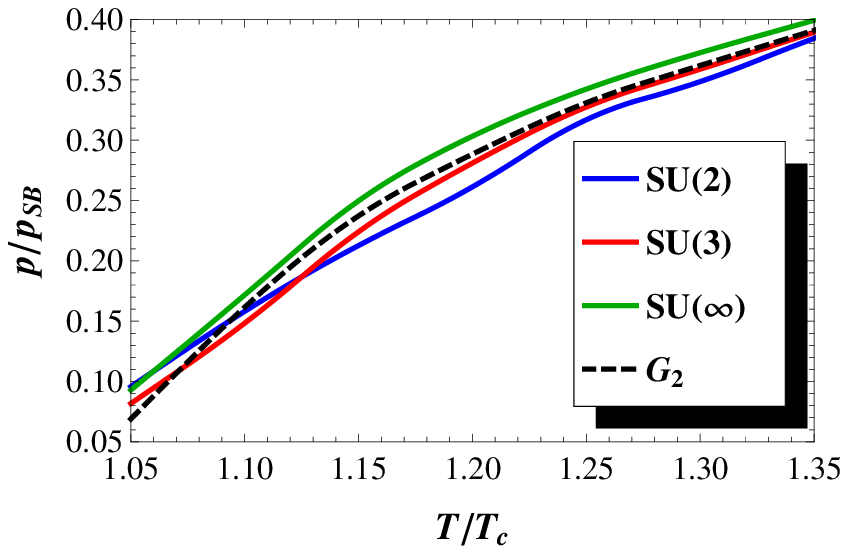}
\caption{Normalized pressure versus temperature in units of $T_c$ (with $T_c$ = 0.3~GeV), computed for the gauge groups SU(2), SU(3), $G_2$, and SU($\infty$). The temperature range is the one where the differences between the curves are the most important. 
}
\label{fig5}
\end{center}
\end{figure}

Notice that our fit of the screening mass does not follow the constraints (\ref{limmu}), but it is designed to fit the static potential below 3~$T_c$. A more involved form would be needed to reach the HTL predictions at high temperatures, but it is not the scope of the present work.

\par In Fig.~\ref{fig4}, the normalized  pressure $P/P_{SB}$ is presented for different gauge groups: SU(2), SU(3), SU($\infty$) and G$_2$. 
Severals remarks can be done. First, the free gluon thermodynamic contribution is gauge-group invariant once normalized to $p_{SB}$. The gauge-group dependence is only present in the bound state and scattering sectors. The number of allowed color channels (\textit{i.e.} the symmetric ones) depends on the gauge group (see (\ref{colo})) and determines the allowed maximum number of bound states and the number of scattering channels. The bound state thermodynamic contribution comes from two effets: The number and the mass of the existing glueballs. Because of the glueball dissociation, this contribution is only taken into account up to the temperature of dissociation (see Tables~\ref{tab:bsatc}). One can observe on Fig.~\ref{fig4} that the produced EoS are not very sensitive to the gauge-group. The most important difference between the curves occurs between 1.05 and 1.35 $T_c$ (see Fig.~\ref{fig5}): In this range, the gluon-gluon interactions are maximal. When the temperature increases, the Born approximation becomes more and more valid and the pressure then scales as $\text{dim}~adj$. Thus the normalized pressure tends to be universal. 

\par In Fig~\ref{fig4}, it is also worth noticing that the EoS computed in our
approach favorably compares with QCD lattice data for gauge groups SU(3-8)
\cite{panero} where such universal curves seem to appear (note that lattice data exist also for very high values of $T/T_c$ but only for SU(3) \cite{bors12}). Concerning G$_2$, no lattice data about EoS are currently available but a
new effective matrix model describing pure Yang-Mills thermodynamics has been
proposed in \cite{dumi}.  These last results are compared to ours in
Fig~\ref{fig4}.

\section{Conclusions}

The relevance of gluon-gluon interactions beyond the critical temperature in the pure gauge
SU(3) plasma has been addressed in a non-perturbative $T$-matrix many-body framework with the
input of Casimir-scaled potentials from thermal lattice QCD and a model of quasigluon mass independent of the gauge group. Scalar glueball bound states in the
singlet channel survive up to temperatures of about 1.3-1.5~$T_c$, together with sizable threshold effects due to strong correlations beyond the two-particle threshold. With only one free parameter, the gluon mass at $T=0$, the EoS of the gluon-glueball gas is reproduced in good agreement with quenched lattice SU($N$) simulations (the other parameters can be fixed \textit{a priori} by resorting to either lattice results or theoretical arguments). Predictions for the $G_2$ EoS are also given, the main feature being that it should be very close to the SU(N) one once normalized to the Stefan-Boltzmann pressure. 

\par The present $T$-matrix formalism can in principle be systematically
improved, for example by including three-gluon scattering. The number of
channels under consideration is quite larger compared with the two-gluon case,
with more elaborated symmetries. Moreover, finding the $T$-matrix would then
become a three-body problem, whose resolution through \textit{e.g.} Faddeev equations can be addressed in future works. 
\par Another natural extension to this paper is to study the light meson spectrum at finite temperature and the QCD EoS  by including quarks within the model. Computations with baryonic potential can be also considered. 
\par Finally, the $T$-matrix formalism can also be applied to calculate bulk thermodynamical properties of the system such as the sheer viscosity, which can be easily computed in relaxation-time approximation within a quasi-particle picture. Such a work is in progress.

\acknowledgments
This work has been partly supported by grant FPA2011-27853-C02-02 (Ministerio de
Econom\'{\i}a y Competitividad, Spain).
D.C. acknowledges financial support from Centro Nacional de F\'{\i}sica de
Part\'{\i}culas, Astropart\'{\i}culas y Nuclear (CPAN, Consolider - Ingenio
2010). G.L. and C.S. thank F.R.S-FNRS for financal supports. The authors thank M. Panero for useful comments.

\begin{appendix}
\section{Some SU($N$) relations}\label{sun}

The dimensions of the color channels appearing in (\ref{colo}) are given in Table~\ref{tab1}, together with the color factors (\ref{kappa0b}). Note that a general method for computing the quadratic Casimir of SU($N$) can be found in \cite{cas}. 

\begin{table}[ht]\caption{Symmetry, dimension (${\rm dim}\ {\cal C}$) and color factor ($\kappa_{{\cal C}}$) defined in (\ref{kappa0b}), of the color channels (${\cal C}$) appearing in the tensor product of the SU($N$) adjoint representation by itself. This table actually displays the two-gluons color channels, denoted by $gg$. The SU(3) case is also indicated.}

\begin{tabular}{c|ccccc}
            
${\cal C}$  for $gg$  & $\bullet$ & $(1,0,\dots,0,1)$ & $(2,0,\dots,0,2)$ & $(2,0,\dots,1,0)$    & $(0,1,0,\dots,0,1,0)$ \\
&             &                 &                 & $(0,1,\dots,2)$  &                     \\
SU(3) & $\bullet$ & (1,1) & (2,2) & (0,3), (3,0) & - \\
\hline
Symmetry & S & S, A & S & A & S  \\
${\rm dim}\ {\cal C}$ & 1     & $N^ 2-1$          &$\frac{N^2(N+3)(N-1)}{4}$ & $\frac{(N^ 2-4)(N^ 2-1)}{4}$ & $\frac{N^2(N-3)(N+1)}{4}$ \\     
$\kappa_{{\cal C}}$ & $-1$    & $-\frac{1}{2}$ & $\frac{1}{N}$ & 0 & $-\frac{1}{N}$ \\

\end{tabular}

\label{tab1}
\end{table}

Similar results can be written in the case where only the fundamental and/or conjugate representations are taken into account. The terms appearing in the tensor product of the fundamental and conjugate representations by themselves are given in Table~\ref{tab3}, as well as those appearing in the tensor product of the fundamental representation by the conjugate one. 

\begin{table}[ht]
\caption{Symmetry, dimension (${\rm dim}\ {\cal C}$) and color factor ($\kappa_{{\cal C}}$) defined in (\ref{kappa0}), of the color channels (${\cal C}$) appearing in the tensor product of the SU($N$) fundamental representation  by itself (left), of the conjugate representation by itself (middle), and of the fundamental representation by the conjugate one (right). This table actually displays the quark-quark, antiquark-antiquark and quark-antiquark color channels, denoted by  $qq$, $\bar q \bar q$, and $q\bar q$ respectively. }
\begin{tabular}{c|cc|cc|cc}
${\cal C}$ for $qq$, $\bar q \bar q$, $q\bar q$  & $(2,0,\dots,0)$ & $(0,1,0,\dots,0)$ & $(0,\dots,0,2)$  &     $(0,\dots,0,1,0)$  & $\bullet$ & $(1,0,\dots,0,1)$ \\
\hline
Symmetry & S & A & S & A & &  \\
${\rm dim}\ {\cal C}$ & $\frac{N(N+1)}{2}$    & $\frac{N(N-1)}{2}$    & $\frac{N(N+1)}{2}$    & $\frac{N(N-1)}{2}$ & 1 &$N^2-1$  \\     
$\kappa_{{\cal C}}$ & $\frac{N-1}{2N^2}$    & $-\frac{N+1}{2N^2}$ & $\frac{N-1}{2N^2}$    & $-\frac{N+1}{2N^2}$ & $-\frac{N^2-1}{2N^2}$ & $\frac{1}{2N^2}$ \\

\end{tabular}

\label{tab3}
\end{table}

Finally, useful results concerning the tensor product of the fundamental (conjugate) representation by the adjoint representation are given in Table~\ref{tab4}.  

\begin{table}[ht]
\caption{Dimension (${\rm dim}\ {\cal C}$) and color factor ($\kappa_{{\cal C}}$) defined in (\ref{kappa0}) of the color channels (${\cal C}$) appearing in the tensor product of the SU($N$) fundamental representation by the adjoint one (left), and of the conjugate representation by the adjoint one (right). This table actually displays the quark-gluon and antiquark-gluon color channels, denoted by $qg$ and $\bar q g$ respectively.}
\begin{tabular}{c|ccc|ccc}
${\cal C}$  for $qg$, $\bar q g$  & $(1,0,\dots,0)$ & $(2,0,\dots,0,1)$ & $(0,1,\dots,0,1)$   & $(0,\dots,0,1)$ & $(1,0,\dots,0,2)$ & $(1,0,...1,0) $\\
\hline
${\rm dim}\ {\cal C}$ & $N$    & $\frac{(N+2)N(N-1)}{2}$    & $\frac{(N+1)N(N-2)}{2}$    & $N$    & $\frac{(N+2)N(N-1)}{2}$    & $\frac{(N+1)N(N-2)}{2}$  \\     
$\kappa_{{\cal C}}$ & $-\frac{1}{2}$    & $\frac{1}{2N}$ & $-\frac{1}{2N}$    & $-\frac{1}{2}$    & $\frac{1}{2N}$ & $-\frac{1}{2N}$  \\

\end{tabular}

\label{tab4}
\end{table}

\section{Lattice potential}\label{appU1}

The lattice data that are used as a starting point to build our interaction potential are those of \cite{kacz3}, \textit{i.e.} the static free energy between a quark-antiquark pair bound in a color singlet for $N_{lat}=3$. For numerical convenience, it is preferable to deal with a fitted form of these, rather than with interpolations of the available points. To fit the data of~\cite{kacz3}, the analytic form proposed by Satz in~\cite{satz} is used:
\begin{equation}\label{f1lat}
F_1(r,T)=\frac{\sigma}{\mu (T)}\left[ \frac{\Gamma(1/4)}{2^{3/2}\Gamma(3/4)}-\frac{\sqrt{\mu(T)r}}{2^{3/4}\Gamma(3/4)}K_{1/4}\left(\mu(T)^2 r^2\right)\right]-\frac{4}{3}\frac{\alpha}{r}\Big[{\rm e}^{-\mu(T) r}+\mu(T) r\Big].
\end{equation}
The way of obtaining this formula is the following. First, it is known that the static quark-antiquark energy at zero temperature is accurately fitted by a so-called funnel shape
\begin{equation}\label{f1z}
F_1(r,0)=\sigma\, r-\displaystyle\frac{4}{3} \displaystyle\frac{\alpha}{r}=U_1(r,0), 
\end{equation}
see \textit{e.g.}~\cite{bali}. When $T>0$, one can imagine that this potential is progressively screened by thermal fluctuations. An effective theory for studying the screening of a given potential is the Debye-H\"uckel theory, in which the thermal fluctuations are all contained in a screening function $\mu(T)$, that modifies the zero-temperature potential and eventually leads to the form (\ref{f1lat}). 

\begin{figure}[ht]
\includegraphics*[width=8.5cm]{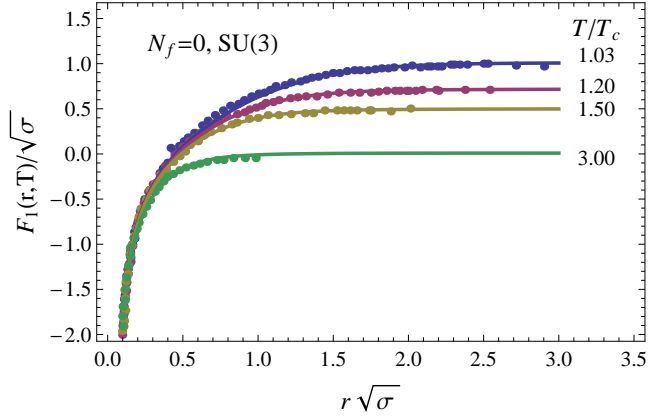}
\caption{Static free energy $F_1(r,T)$ of a quark-antiquark pair bound in a color singlet, computed in SU(3) quenched lattice QCD and plotted for different temperatures (symbols). Data are taken from~\cite{kacz3} and expressed in units of $\sqrt\sigma$, with $r$ the quark-antiquark separation. The fitted form (\ref{f1lat})-(\ref{afit}) is compared to the lattice data (solid lines). }
\label{fig1a}
\end{figure}

\begin{figure}[ht]
\includegraphics*[width=8.5cm]{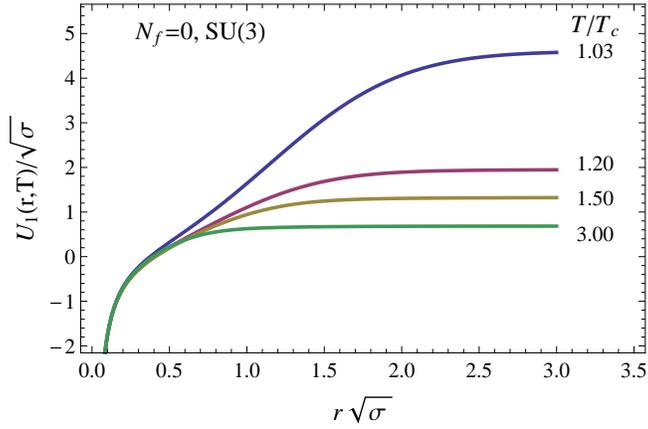}
\caption{Internal energy $U_1(rT)$ of a quark-antiquark pair bound in a color singlet, computed from the fitted form (\ref{f1lat})-(\ref{afit}) and plotted for different temperatures (solid lines). }
\label{fig2a}
\end{figure}

The explicit form of $\mu(T)$ is unknown \textit{a priori} and has to be fitted on the lattice data. As it can be seen in Fig.~\ref{fig1a}, the form
\begin{eqnarray}\label{mufit}
\frac{\mu(T)}{\sqrt\sigma}&=&0.537\frac{T}{T_c}+0.644+0.112\ln\left(\frac{T}{T_c}-
0.967\right),
\end{eqnarray}
with
\begin{equation}\label{afit}
\alpha=0.141,
\end{equation}
provides an accurate fit of the lattice data in the range 1-3~$T_c$. A more complete fit should be such that $\mu(0)=0$, but our model is not intended to be able to ``cross" the phase transition in $T_c$. The simple form (\ref{mufit}) is already satisfactory. The corresponding internal energy $U_1=F_1-T\partial_T F_1$ is plotted in Fig.~\ref{fig2a}.

\section{Dashen, Ma and Bernstein's formalism in momentum space} \label{trace}

To compute (\ref{omega2}), it is necessary to use a given representation. In order to use the calculation of the $T$-matrix proposed in \cite{cabre06}, the scattering part of (\ref{pot_s3}) must be computed in the momentum space representation (the two first term are simply free gas contributions and can be easily computed). Let us focus on 
\begin{eqnarray} \label{omS}
\Omega_s &=& \sum_{{\cal C}}\sum_{J^{P}}\frac{{\rm dim}\, {\cal C}}{2\pi^2\beta^2}\, (2J+1)\int^\infty_{2m_g} d\epsilon \, \epsilon^2\,  K_2(\beta\epsilon)  \nonumber \\   
&& \times{\rm Tr}_{{\cal C},J^{P}} \Bigg[ \left( \delta {\rm Re} {\cal T }\right)'- 2\pi \Big( (\delta {\rm Re}{\cal  T})(\delta {\rm Im}{\cal T})'-(\delta {\rm Im} {\cal  T}) (\delta {\rm Re}{\cal  T})'  \Big) \Bigg] . 
\end{eqnarray}
Using the following definitions concerning the trace of an operator $A$ in momentum space
\begin{equation}
{\rm Tr} A = \displaystyle\frac{1}{(2\pi)^ 3} \displaystyle\int_{-\infty}^{\infty} d\vec{q} \, \langle \vec{q}\, | A | \vec{q} \,\rangle
\end{equation}
and the partial wave expansion
\begin{equation}
\langle \vec{q}\, | A | \vec{q} \,'\rangle = A(q,q',\hat{q}.\hat{q}') = \displaystyle\frac{1}{4\pi} \displaystyle\sum_l (2l + 1) A_l(q,q') P_l(\hat{q}.\hat{q}') 
\end{equation}
where $P_l(x)$ is the Legendre polynomial of order $l$, (\ref{omS}) reads
\begin{eqnarray}
\Omega_s &=& \displaystyle\frac{1}{64 \pi^ 5\beta^ 2} \displaystyle\sum_{J^P} (2 J + 1) \displaystyle\sum_{{\cal C}} \text{dim} {\cal C} \left(  \beta \displaystyle\int_{2m_g}^\infty d\epsilon \, \epsilon^3 \displaystyle\sqrt{\frac{\epsilon^2}{4} - m_g^2} K_1(\beta\epsilon) \,{\rm Re} {\cal T}_{J^P}(\epsilon; q_\epsilon, q_\epsilon) \right.\nonumber \\ \nonumber
&& -\left. \displaystyle\frac{1}{16 \pi^2} \displaystyle\int_{2m_g}^\infty d\epsilon \, \epsilon^4 \, \left(\frac{\epsilon^2}{4} - m_g^2 \right) K_2(\beta\epsilon) \left[{\rm Re} {\cal T}_{J^P}(\epsilon; q_\epsilon, q_\epsilon) \left({\rm Im} {\cal T}_{J^P}(\epsilon; q_\epsilon, q_\epsilon) \right)' \right]  \right. \\ 
&& + \left. \displaystyle\frac{1}{16 \pi^2} \displaystyle\int_{2m_g}^\infty d\epsilon \, \epsilon^4 \, \left(\frac{\epsilon^2}{4} - m_g^2\right) K_2(\beta\epsilon) \left[\left({\rm Re} {\cal T}_{J^P}(\epsilon; q_\epsilon, q_\epsilon)\right)' {\rm Im} {\cal T}_{J^P}(\epsilon; q_\epsilon, q_\epsilon)  \right]  \right).
\end{eqnarray}
\end{appendix}

\end{document}